\pdfoutput=1
\documentclass[journal]{IEEEtran}
\usepackage{cite}
\usepackage{graphicx}
\usepackage{amsmath}
\usepackage{amssymb}
\usepackage{comment}
\usepackage{booktabs}
\usepackage{booktabs}
\usepackage{xcolor}
\usepackage{booktabs}
\usepackage{siunitx}
\usepackage{algorithm}
\usepackage{float}
\usepackage{algpseudocode}
\usepackage{placeins}
\usepackage{tikz}
\usetikzlibrary{shapes.geometric, arrows.meta, positioning, fit, backgrounds, calc, shadows}

\definecolor{blueFill}{RGB}{221,235,250}
\definecolor{blueLine}{RGB}{58,111,163}
\definecolor{grayFill}{RGB}{234,234,236}
\definecolor{grayLine}{RGB}{110,110,116}
\definecolor{pinkFill}{RGB}{250,222,219}
\definecolor{pinkLine}{RGB}{178,60,55}
\definecolor{greenFill}{RGB}{222,238,217}
\definecolor{greenLine}{RGB}{74,124,64}
\definecolor{zoneRed}{RGB}{178,60,55}
\definecolor{zoneOrange}{RGB}{198,124,42}
\definecolor{zoneBlue}{RGB}{40,80,130}

\allowdisplaybreaks

\IEEEoverridecommandlockouts
\IEEEoverridecommandlockouts
\IEEEpubid{\footnotesize This work has been submitted to the IEEE for possible publication. Copyright may be transferred without notice, after which this version may no longer be accessible.}

\begin{document}
\raggedbottom

\title{Grey-Box Experimental Identification of Reduced-Order Models for Grid-Forming Converters}

\author{Marcelo~Cardenas,~\IEEEmembership{Student~Member,~IEEE,}
Stephan~Kenzelmann, Matteo~Saviozzi,~\IEEEmembership{Senior~Member,~IEEE,} Federico~Silvestro,~\IEEEmembership{Senior~Member,~IEEE,} and  Fabrizio~Sossan}
\maketitle

\begin{abstract}
This paper proposes a grey-box system identification for identifying and validating reduced-order models (ROMs) of grid-forming (GFM) converters. Three reduced-order modeling techniques are compared within a common identification procedure: singular perturbation with frequency cutoff and participation factors (SP-PF), singular perturbation with energy residuals (SP-Energy), and balanced truncation (BT). Testing is on experimental data with real-life components. Results show that BT achieves the highest time-domain fit, followed by SP-Energy and SP-PF;
SP-Energy better preserves key properties of the identified full-order model; BT places the dominant mode closest to the stability boundary and exhibits the highest sensitivity to parameter variations; SP-Energy outperforms SP-PF in both fit quality and consistency of modal preservation. Findings suggest that the model with the best time-domain fit does not always provide the most reliable representation of the modal properties for small-signal stability, with SP-Energy offering the best compromise between accuracy, physical interpretability, and modal preservation for the considered case study.

\end{abstract}


\section{Introduction}
\IEEEPARstart{T}{he} transition from synchronous machines (SMs) to converter-interfaced resources is fundamentally redefining power system dynamics. Grid Forming (GFM) converters are becoming critical for system stability thanks to providing voltage, frequency regulation, and virtual inertia support \cite{qi2022small, li2024modelling,11143695}. Reliable models of GFM converters are essential for accurate stability studies; however, models adopted for such studies often trade accuracy for computational tractability. E.g., while inertia-placement studies in \cite{gotti2023recursive} rely on second-order angle–inertia models that neglect converter control dynamics, detailed converter models generally exhibit orders exceeding 15 due to the complexity of their internal control and electrical dynamics \cite{wang2023small}.

Reduced-order modeling (ROM) techniques such as singular perturbation (SP) and balanced truncation (BT) have been explored to simplify models while retaining the dominant dynamics (e.g., \cite{CADUFF2021106773}). Grey-box identification methods have also been proposed to estimate key parameters, including inertia and droop coefficients, from simulation data using linear time-invariant representations that capture power and voltage dynamics, but do not fully represent the full-order model (FOM) of the converter or its interaction with the grid \cite{10842804,10485402,9990926}.

Parameter estimation in Inverter Based Resources (IBR)-dominated power systems are addressed in several perspectives. At the converter level, real time estimation methods for GFM converters have mainly focused on external grid parameters, such as grid impedance, while other works have investigated inertial response estimation and parameter tuning based on a operating region that introduces a stability factor \cite{zhao2024real,10856112}. While these approaches provide useful insight into adaptive frequency support, they do not address converter level grey-box identification from experimental measured data.

A major modeling challenge is that commercial GFM converters are black-boxes: internal control structures and parameters are not disclosed due to intellectual property restrictions, limiting the applicability of purely physics based white-box modeling in system studies \cite{10842804,11221946}. This motivates grey-box identification, which combines measured input-output data with a structure to preserve model formulation to estimate  parameters \cite{10842804}. Grey-box identification is particularly suitable for GFM converters because it links measured dynamic responses to different parameters associated with the GFM formulation (e.g., inertia, droop, damping related coefficients and PI gains) \cite{9451617}. Despite this, the literature still lacks an experimentally validated framework based on a structured and physically model representation that preserves key states, compares full-order and multiple reduced-order realizations, and evaluates how identification accuracy relates to modal stability preservation within a single workflow. This paper aims to bridge this gap by proposing a framework for estimating and validating ROMs from measurements to retain tractability and enhance adaptability.

The main contributions of this paper are summarized as follows:
\begin{itemize}
\item A grey-box identification framework that combines experimental measured input--output responses of a grid-connected GFM converter with a structured model representation, enabling both FOM and ROM 
identification within a single parameterized realization.
\item A constrained parameter estimation workflow that estimates a selected subset of physically meaningful degrees of freedom (e.g., inertia-, droop-, and damping related parameters, as well as PI gains) under stability screening.

\item A systematic comparison of ROM techniques, including Singular Perturbation with Frequency Cutoff and Participation Factors (SP-PF), Balanced Truncation (BT), and, the energy-based SP variant introduced within the proposed framework. In particular, SP-Energy extends SP-PF by replacing the frequency-only slow/fast partition with a residue-based energy criterion.
\end{itemize}

The remainder of the paper is organized as follows.
Section~II presents the problem statement and notation.
Section~III presents the formulation of the three ROMs.
Section~IV presents the grey-box modeling procedure and the performance metrics.
Section~V presents the nominal FOM and the converter architecture.
Section~VI describes the experimental setup and the dataset.
Section~VII reports the results and discusses the trade-off between time-domain fit, modal preservation, and physical interpretability.
Finally, Section~VIII concludes the paper.


\section{Problem Statement and Notation}
\label{sec:problem}

The objective is to develop and identify ROMs for power converters and to evaluate their performance. We adopt a grey-box modeling
strategy, consisting of the following steps:

\begin{itemize}
    \item candidate model structures are identified based on physical system
    knowledge or rigorously justified modeling assumptions;
    \item their parameters are estimated from experimental measurements; and
    \item model performance is assessed and compared in terms of time-domain
    response, preservation of the dominant eigenvalues, damping characteristics,
    stability margins, and the physical interpretability of the identified parameters.
\end{itemize}

The considered power converter topology is shown in Fig.~\ref{fig:gfm1}(a). It consists of the physical power conversion stage (switching
devices, filter and grid connection) and its GFM control structure (active and
reactive power controllers, cascaded voltage and current control loops, and the transient virtual impedance).

\subsection{Notation}
\label{subsec:notation_problem}

Throughout the paper, \(x\in\mathbb{R}^{n}\), \(u\in\mathbb{R}^{m}\), and
\(y\in\mathbb{R}^{p}\) denote the state, input, and output vectors of a
continuous-time linear model, and \(\theta\in\mathbb{R}^{q}\) collects the
physical and control parameters of the converter. Small-signal quantities are
denoted by \(\Delta(\cdot)\); the operator is omitted when the linear nature of
the model is clear from the context. The angle \(\delta\) between the Inverter Reference Frame (IRF) and the Grid Reference Frame (GRF) is used throughout Section~\ref{sec:gfm_full}. Subscripts \(d\) and \(q\) denote the direct- and quadrature-axis components of a \(dq\)-frame vector, respectively. The parameter symbols used for
the converter (virtual inertia \(H_v\), droop \(R_p\), damping \(D_p\), PI gains,
feedforward gains, and virtual impedance) are listed in
Table~\ref{tab:params_design_control}.

\subsection{Full-Order Model Notation}
\label{subsec:problem_fom}

The behavior of the power converter is described by a small-signal model around an
operating point that reads as:
\begin{equation}
\begin{aligned}
\dot{x}(t) &= A(\theta)x(t) + B(\theta)u(t),\\
y(t) &= C(\theta)x(t) + D(\theta)u(t),
\end{aligned}
\label{eq:prob_fom}
\end{equation}
For the converter topology under study, the system order is \(n=19\). Section~\ref{sec:gfm_full} details the state definitions, subsystem interconnections, and the CCM-based assembly of this realization.

\subsection{Reduced-Order Model Notation}
\label{subsec:problem_rom}

This paper is concerned with finding reduced-order models (ROMs) of order \(r<n\),
for improved tractability of large-scale power grid studies. ROMs are derived from
the FOM and share its parameterization \(\theta\). Their structure is
\begin{equation}
\begin{aligned}
\dot{x}_r(t) &= A_r(\theta)x_r(t) + B_r(\theta)u(t),\\
y_r(t) &= C_r(\theta)x_r(t) + D_r(\theta)u(t).
\end{aligned}
\label{eq:prob_rom}
\end{equation}

Three ROM candidate structures are proposed and explored in this paper:
\begin{itemize}
    \item singular perturbation with frequency cutoff and participation factors (SP-PF),
    \item singular perturbation with an energy-based modal criterion (SP-Energy),
    \item balanced truncation (BT).
\end{itemize}
Together with the FOM, they define the set of model structures
\(M\in\{\mathrm{Full},\,\mathrm{SP\!-\!PF},\,\mathrm{SP\!-\!Energy},\,\mathrm{BT}\}\)
considered throughout the paper, which differ only in the criterion used to reduce
the FOM dynamics. In the results, the structure \(M=\mathrm{Full}\) evaluated at the
parameter values of Table~\ref{tab:params_design_control} is referred to as the
nominal model, and Full denotes the same structure after parameter estimation. Given experimental input--output measurements from the converter, the problem is to
estimate the parameter vector \(\theta\) for each candidate structure, and to assess
how the resulting models reproduce the measured response and preserve the dynamic
properties of the FOM from which they are derived. This is formulated in
Section~\ref{sec:greybox}.

\section{Formulation of Reduced-Order Models}
\label{sec:roms}

This section presents the three reduced-order modeling strategies considered in this
work. All of them are applied to the full-order realization \((A,B,C,D)\) of
\eqref{eq:prob_fom}, and each of them produces a reduced realization
\((A_r,B_r,C_r,D_r)\) in the form of \eqref{eq:prob_rom}. All reductions are applied to internally stable realizations satisfying the stability
condition in Section~\ref{subsec:gb_problem}. The numerical values adopted for
these hyperparameters are $f_{\mathrm{cut}}=1000$~Hz (SP cutoff frequency),
$p_{\mathrm{th}}=0.7$ (SP-Energy threshold), $[0,f_{\max}]=10^6$~Hz
(SP-Energy frequency range), and $\eta^\star=99.95\%$ (BT HSV threshold).

\subsection{Singular Perturbation with Frequency Cutoff and Participation Factors (SP-PF)}
\label{subsec:sp_pf}

The SP-PF approach identifies slow and fast system modes using a frequency cutoff and then uses participation factors to determine which states are mainly associated with each group. The fast states are subsequently eliminated using singular perturbation, yielding a ROM that preserves the relevant slow dynamics \cite{kimball2008singular,rasheduzzaman2015reduced,WANG2020106030,CADUFF2021106773}.

\(\lambda_k\) denotes the eigenvalues of \(A\), with associated right eigenvectors
\(v_k\) satisfying \(A v_k=\lambda_k v_k\). The modal frequency and damping ratio of
mode \(k\) are
\begin{equation}
\omega_k = |\Im(\lambda_k)|,\qquad
f_k=\frac{\omega_k}{2\pi},\qquad
\zeta_k=-\frac{\Re(\lambda_k)}{|\lambda_k|}.
\label{eq:rom_fk_zetak}
\end{equation}
Using the cutoff \(f_{\mathrm{cut}}\), the modes are separated into slow
\((\mathcal{S})\) and fast \((\mathcal{F})\) sets:
\begin{equation}
\mathcal{S}=\{k:\, f_k \le f_{\mathrm{cut}}\}, \qquad
\mathcal{F}=\{k:\, f_k > f_{\mathrm{cut}}\}.
\label{eq:rom_SF_sets}
\end{equation}

Let \(V=[v_1,\dots,v_n]\) and \(W=[w_1,\dots,w_n]\) denote the matrices of right and
left eigenvectors. The participation factor
of state \(i\) in mode \(k\) is
\begin{equation}
PF_{ik}=\left|v_{ik} w_{ik}\right|,
\qquad
\widetilde{PF}_{ik}=\frac{PF_{ik}}{\max_i PF_{ik}},
\label{eq:rom_pf_def}
\end{equation}
where \(\widetilde{PF}_{ik}\) is column-normalized. The aggregated slow and fast
relevance of state \(x_i\) is defined as
\begin{equation}
PF^{(s)}_i = \frac{1}{|\mathcal{S}|}\sum_{k\in\mathcal{S}} \widetilde{PF}_{ik}, \qquad
PF^{(f)}_i = \frac{1}{|\mathcal{F}|}\sum_{k\in\mathcal{F}} \widetilde{PF}_{ik}.
\label{eq:rom_pf_agg}
\end{equation}
State \(x_i\) is classified as slow if \(PF^{(s)}_i \ge PF^{(f)}_i\), and as fast
otherwise.

After reordering the state vector as \(x=[x_s^\top\;x_f^\top]^\top\), the system
matrices are written as
\begin{equation}
A=
\begin{bmatrix}
A_{ss} & A_{sf}\\
A_{fs} & A_{ff}
\end{bmatrix},\quad
B=
\begin{bmatrix}
B_s\\
B_f
\end{bmatrix},\quad
C=
\begin{bmatrix}
C_s & C_f
\end{bmatrix}.
\label{eq:rom_blocks}
\end{equation}
The fast states are approximated by their quasi-steady-state value
\begin{equation}
x_f = -A_{ff}^{-1}(A_{fs}x_s + B_f u).
\label{eq:rom_sp_xf}
\end{equation}
Only the matrices \(A\), \(B\), and \(C\) are decomposed, whereas \(D\) remains
unchanged with respect to the CCM stage. The resulting reduced realization, stated
in \cite{rasheduzzaman2015reduced}, is
\begin{equation}
\begin{aligned}
A_{\mathrm{SP}} &= A_{ss}-A_{sf}A_{ff}^{-1}A_{fs},\\
B_{\mathrm{SP}} &= B_s-A_{sf}A_{ff}^{-1}B_f,\\
C_{\mathrm{SP}} &= C_s-C_fA_{ff}^{-1}A_{fs},\\
D_{\mathrm{SP}} &= D-C_fA_{ff}^{-1}B_f,
\end{aligned}
\label{eq:rom_sp_mats}
\end{equation}
and the complete procedure is summarized in Algorithm~\ref{alg:sp_pf}.

\begin{algorithm}[!t]
\footnotesize
\caption{SP-PF Reduced-Order Model Construction}
\label{alg:sp_pf}
\begin{algorithmic}[1]
\Require Full-order realization \((A,B,C,D)\) and cutoff frequency \(f_{\mathrm{cut}}\)
\Ensure Reduced realization \((A_{\mathrm{SP}},B_{\mathrm{SP}},C_{\mathrm{SP}},D_{\mathrm{SP}})\)
\State Compute eigenpairs $(\lambda_k,v_k)$ and left eigenvectors $w_k$ of $A$, normalized such that $w_k^\top v_k=1$
\State Compute the modal frequencies \(f_k\) and classify the modes into \(\mathcal{S}\) and \(\mathcal{F}\) using \eqref{eq:rom_fk_zetak}--\eqref{eq:rom_SF_sets}
\State Compute \(\widetilde{PF}_{ik}\) using \eqref{eq:rom_pf_def}
\State Compute the aggregated slow and fast state relevance using \eqref{eq:rom_pf_agg}
\State Classify the states into \(x_s\) and \(x_f\)
\State Form block matrices $(A_{ss},A_{sf},A_{fs},A_{ff},B_s,B_f,C_s,C_f)$ as in \eqref{eq:rom_blocks}
\State Apply \eqref{eq:rom_sp_xf}--\eqref{eq:rom_sp_mats} to obtain \((A_{\mathrm{SP}},B_{\mathrm{SP}},C_{\mathrm{SP}},D_{\mathrm{SP}})\)
\end{algorithmic}
\end{algorithm}

\subsection{Singular Perturbation with Modal Energy Criterion (SP-Energy)}
\label{subsec:sp_energy}

The SP-Energy approach identifies slow and fast system modes using a residue-based modal energy metric and then uses participation factors to determine which states are mainly associated with each group, keeping the reduction step unchanged. In SP-PF, the slow/fast partition is defined by a frequency-based modal criterion; here it is redefined through modal energy \cite{GAROFALO2002125,VASSILYEV201714806}. Energy-based modal indicators have been used mainly for stability assessment \cite{ISKAKOV2021109814}, rather than for constructing the slow/fast partition for singular-perturbation reduction. To the best of the authors' knowledge, this use has not been reported.

Let \((\lambda_k,v_k)\) and \(w_k\) denote the right and left eigenvectors of mode
\(k\). The residue matrix of mode \(k\) is
\begin{align}
R_k &= C v_k w_k^\top B, \label{eq:rom_residue}\\
G_k(j\omega)&=\frac{R_k}{j\omega-\lambda_k},
\qquad
g_k(\omega)=\|G_k(j\omega)\|_F, \label{eq:rom_Gk}
\end{align}
where \(g_k(\omega)\) quantifies the input--output contribution of mode \(k\) at
frequency \(\omega\) \cite{GAROFALO2002125}. Let \(\omega_c = 2\pi f_{\mathrm{cut}}\)
and \(\omega_{\max}=2\pi f_{\max}\), where \(f_{\max}\) is the upper limit of the
frequency grid used in the energy calculation. The high-frequency energy fraction of
mode \(k\) is defined as
\begin{equation}
\phi_k=
\frac{\int_{\omega_c}^{\omega_{\max}} g_k(\omega)^2\,\mathrm{d}\omega}
{\int_{0}^{\omega_{\max}} g_k(\omega)^2\,\mathrm{d}\omega}.
\label{eq:rom_phi}
\end{equation}
Thus, \(\phi_k\in[0,1]\) measures the fraction of the modal contribution concentrated above the cut-off frequency. Here, \(p_{\mathrm{th}} \in (0,1)\) denotes the threshold used for modal energy classification. Modes are classified as
\begin{equation}
\mathcal{F}=\{k:\phi_k>p_{\mathrm{th}}\},\qquad
\mathcal{S}=\{k:\phi_k\le p_{\mathrm{th}}\}.
\label{eq:rom_phi_sets}
\end{equation}

Using the normalized participation factors \(\widetilde{PF}_{ik}\) from
\eqref{eq:rom_pf_def}, define
\begin{equation}
PF^{(s)}_i=\sum_{k\in\mathcal{S}} \widetilde{PF}_{ik}(1-\phi_k),\qquad
PF^{(f)}_i=\sum_{k\in\mathcal{F}} \widetilde{PF}_{ik}\phi_k.
\label{eq:rom_pf_energy}
\end{equation}
State \(x_i\) is assigned to the slow subset if \(PF^{(s)}_i \ge PF^{(f)}_i\), and to
the fast subset otherwise. The reduced realization
\((A_{\mathrm{SPe}},B_{\mathrm{SPe}},C_{\mathrm{SPe}},D_{\mathrm{SPe}})\) is then
obtained from \eqref{eq:rom_sp_xf}--\eqref{eq:rom_sp_mats}.
Algorithm~\ref{alg:sp_energy} summarizes the procedure.

\begin{algorithm}[!t]
\footnotesize
\caption{SP-Energy Reduced-Order Model Construction}
\label{alg:sp_energy}
\begin{algorithmic}[1]
\Require Full-order realization \((A,B,C,D)\), cutoff \(f_{\mathrm{cut}}\), threshold \(p_{\mathrm{th}}\), and frequency range \([0,f_{\max}]\)
\Ensure Reduced realization \((A_{\mathrm{SPe}},B_{\mathrm{SPe}},C_{\mathrm{SPe}},D_{\mathrm{SPe}})\)
\State Compute the eigenpairs \((\lambda_k,v_k)\) of \(A\) and the corresponding left eigenvectors \(w_k\), normalized such that \(w_k^\top v_k=1\)
\State Compute the modal residues \(R_k\) using \eqref{eq:rom_residue}
\State Compute \(\phi_k\) from \eqref{eq:rom_phi}
\State Classify the modes into \(\mathcal{S}\) and \(\mathcal{F}\) using \eqref{eq:rom_phi_sets}
\State Compute \(PF^{(s)}_i\) and \(PF^{(f)}_i\) using \eqref{eq:rom_pf_energy}, and partition the states into \(x_s\) and \(x_f\)
\State Form the block matrices \((A_{ss},A_{sf},A_{fs},A_{ff},B_s,B_f,C_s,C_f)\) as in \eqref{eq:rom_blocks}
\State Apply the SP reduction \eqref{eq:rom_sp_xf}--\eqref{eq:rom_sp_mats} to obtain \((A_{\mathrm{SPe}},B_{\mathrm{SPe}},C_{\mathrm{SPe}},D_{\mathrm{SPe}})\)
\end{algorithmic}
\end{algorithm}

\subsection{Balanced Truncation (BT)}
\label{subsec:bt}

Balanced truncation (BT) is based on joint controllability and observability of the system states
\cite{Ghosh15052013,doi:10.1049/iet-gtd.2014.1116,nguyen2025efficient}.

For an internally stable realization, the controllability and observability Gramians
\(P\) and \(Q\) are computed from
\begin{equation}
A P + P A^\top + B B^\top = 0,\qquad
A^\top Q + Q A + C^\top C = 0.
\label{eq:rom_lyap}
\end{equation}
A balancing transformation \(T\) gives the balanced realization
\begin{equation}
A_b=T^{-1}AT,\quad B_b=T^{-1}B,\quad C_b=CT,\quad D_b=D,
\label{eq:rom_balanced}
\end{equation}
where \(A_b,B_b,C_b,D_b\) denote the balanced state-space matrices. In these coordinates, the Gramians are transformed such that
\begin{equation}
P_b=Q_b=\Sigma=
\mathrm{diag}(\sigma_1,\ldots,\sigma_{n_B}),
\qquad
\sigma_1\geq\cdots\geq\sigma_{n_B},
\label{eq:rom_hsv}
\end{equation}
where \(\sigma_i\) are the Hankel singular values (HSVs) and \(n_B \le n\) denotes the order of the minimal realization obtained before balancing. States associated with the smallest HSVs have the weakest contribution to the input--output dynamics and are therefore discarded.

The reduced order \(r\) is selected using the cumulative HSV ratio
\begin{equation}
\eta_r =
\frac{\sum_{k=1}^{r}\sigma_k}
{\sum_{k=1}^{n_B}\sigma_k},
\qquad
r=\min\{r:\eta_r\geq\eta^\star\},
\label{eq:bt_energy_ratio}
\end{equation}
where \(\eta^\star\) is the prescribed threshold.

After partitioning the balanced realization according to the retained and discarded coordinates, the BT model is obtained by retaining the first \(r\) balanced states,
\begin{equation}
A_{\mathrm{BT}}=A_{11},\qquad
B_{\mathrm{BT}}=B_1,\qquad
C_{\mathrm{BT}}=C_1,\qquad
D_{\mathrm{BT}}=D.
\label{eq:rom_bt_rom}
\end{equation}
The associated \(\mathcal{H}_\infty\) error bound is \cite{Ghosh15052013}
\begin{equation}
\|G-G_r\|_\infty \le 2\sum_{i=r+1}^{n_B}\sigma_i.
\label{eq:rom_bt_bound}
\end{equation}
Unlike the SP-based approaches, BT does not preserve individual physical states, since the balanced coordinates are linear combinations of the original states.

\begin{algorithm}[!t]
\footnotesize
\caption{Balanced Truncation Model}
\label{alg:bt_hsv}
\begin{algorithmic}[1]
\Require Full-order realization \((A,B,C,D)\), HSV threshold \(\eta^\star\)
\Ensure BT reduced model \((A_{\mathrm{BT}},B_{\mathrm{BT}},C_{\mathrm{BT}},D_{\mathrm{BT}})\) and order \(r\)
\State Compute the minimal realization \((A_m,B_m,C_m,D_m)\)
\State \(n_B\) is the order of the balanced minimal realization
\State Solve the Lyapunov equations in \eqref{eq:rom_lyap} and compute the balanced realization in \eqref{eq:rom_balanced}
\State Extract the Hankel singular values \(\{\sigma_i\}_{i=1}^{n_B}\)
\State Compute the cumulative ratio \(\eta_r\) using \eqref{eq:bt_energy_ratio}
\State Select \(r=\min\{r:\eta_r\ge\eta^\star\}\) with \(r<n_B\)
\State Construct the BT ROM via truncation using \eqref{eq:rom_bt_rom}
\State Evaluate the \(\mathcal{H}_\infty\) error bound using \eqref{eq:rom_bt_bound}
\end{algorithmic}
\end{algorithm}

\section{Grey-Box Modeling Procedure}
\label{sec:greybox}

\begingroup\sloppy
For each structure \(M\) defined in
Section \ref{subsec:problem_rom}, the corresponding realization is denoted
\((A_M(\theta),B_M(\theta),\) \(C_M(\theta),D_M(\theta))\): for \(M=\mathrm{Full}\) it is
the FOM \eqref{eq:prob_fom}, and otherwise the reduced realization obtained with the
associated reduction method. The same procedure is applied to the four model
structures, so that all of them are compared under identical data, parameterization,
and stability requirements.
\par\endgroup

\subsection{General Procedure}
\label{subsec:workflow}

\begin{figure*}[t!]
      \vspace{-1.1cm}
    \centering
    \includegraphics[trim={0 2 0 0},clip,width=0.925\textwidth]{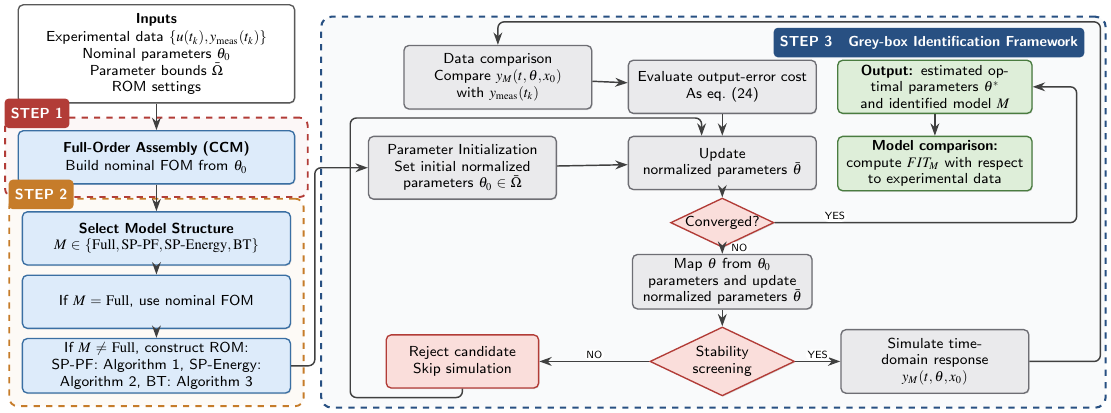}
    \caption{Workflow of the proposed reduced-order modeling and grey-box identification methodology.}
    \label{fig:workflow}
\end{figure*}

Fig.~\ref{fig:workflow} summarizes the proposed workflow. The procedure starts from
the experimental input--output data \(\{u(t_k), y_{\mathrm{meas}}(t_k)\}_{k=1}^{N}\),
the nominal parameter vector \(\theta_0\), the parameter bounds \(\bar{\Omega}\), and
the ROM settings given in Section~\ref{sec:roms}, and consists of three steps.

\paragraph{Step~1: Full-order model assembly}
The nominal FOM \eqref{eq:prob_fom} is assembled from the converter
subsystems as described in Section~\ref{sec:gfm_full}, giving
\((A_{\mathrm{Full}},B_{\mathrm{Full}},C_{\mathrm{Full}},D_{\mathrm{Full}})\) and the
nominal parameter vector \(\theta_0\) of Table~\ref{tab:params_design_control}.

\paragraph{Step~2: Model structure selection}
A model structure \(M\) is selected. If \(M=\mathrm{Full}\), the CCM realization is
used directly; otherwise, the corresponding reduction of Section~\ref{sec:roms} is
applied to the full-order realization.

\paragraph{Step~3: Grey-box identification}
The identification problem is formulated as a bound-constrained nonlinear
least-squares problem. At each iteration, the normalized parameter vector is updated,
the realization of the selected structure is rebuilt, and the candidate is screened
for stability before time-domain simulation. Stable candidates are simulated and
compared with the measured outputs through an output-error cost function. The
procedure continues until convergence and returns the estimated parameter set
\(\theta^\star_M\) and the identified realization.

\subsection{Measurements Used for Identification}
\label{subsec:measurements}

The identification uses a single grid-connected experiment in which the converter is
subjected to a sequence of active power reference steps. Accordingly, the input is
the applied reference variation
\begin{equation}
u(t) = \Delta P_{\mathrm{ref}}(t),
\label{eq:gb_input}
\end{equation}
and the measured outputs are the active and reactive power injected by the converter,
\begin{equation}
y_{\mathrm{meas}}(t) =
\begin{bmatrix}
\Delta P_e(t) & \Delta Q_e(t)
\end{bmatrix}^{\top},
\label{eq:gb_output}
\end{equation}
obtained from the RMS voltage and current measurements acquired at the converter
terminals, as described in Section~\ref{sec:exp}. The remaining inputs of the FOM
(\(\Delta Q_{\mathrm{ref}}\) and the grid voltage components) are held at their
operating-point values during the experiment and are therefore not excited. The
model output \(y_M(t)\) is mapped to the same two channels through
\(C_M(\theta)\) and \(D_M(\theta)\), so that model and measurement are compared on
identical quantities. The signals are sampled at the acquisition rate of the
measurement devices, giving the dataset
\(\{u(t_k),y_{\mathrm{meas}}(t_k)\}_{k=1}^{N}\) used in
\eqref{eq:gb_opt}.

\subsection{Estimation Problem}
\label{subsec:gb_problem}

The parameter vector is partitioned as \(\theta=[\theta_f^\top\ \theta_c^\top]^\top\),
where \(\theta_f\) collects the parameters kept fixed at their nominal values and
\(\theta_c\) the estimated degrees of freedom (DoF). The selected DoF are
\begin{equation}
\begin{aligned}
\theta_c =
\big[
&H_v,\ R_p,\ \beta_I,\ \beta_V,\ K_{ii},\ K_{iq},\ K_{iv},\\
&K_{pi},\ K_{pq},\ K_{pv},\ l_v,\ D_p,\ r_v
\big]^\top .
\end{aligned}
\label{eq:theta_def}
\end{equation}

The relative bounds assigned to the DoF are
$[0.5\,\theta_0,\,2\,\theta_0]$ for $H_v, R_p, K_{pq}, K_{iq}, D_p$;
$[0.5\,\theta_0,\,\theta_0]$ for $\beta_I, \beta_V$; and
$[0.05\,\theta_0,\,5\,\theta_0]$ for $K_{iv}, K_{pi}, K_{pv}, K_{ii}, l_v, r_v$,
where \(\theta_0\) denotes the nominal value of the
corresponding parameter reported in Table~\ref{tab:params_design_control}. All model
structures share this symbolic parameterization; they differ only in the realization
\((A_M,B_M,C_M,D_M)\) used to represent the dynamics.

The estimated parameters are normalized and constrained by fixed lower and upper bounds. Continuous-time stability requires all eigenvalues of \(A_M(\theta)\) to have negative real parts: \(\max_i \Re\{\lambda_i(A_M(\theta))\} < 0\). Candidates that violate this condition are discarded, thereby restricting the search to stable dynamical regions. Given the dataset of Section~\ref{subsec:measurements}, the identification is
formulated as a bound-constrained output-error problem:
\begin{equation}
\bar{\theta}_M^\star=\arg\min_{\bar{\theta}_c\in\bar{\Omega}}
\sum_{k=1}^{N}
\left\|
y_{\mathrm{meas}}(t_k)-y_M(t_k;\theta(\bar{\theta}_c))
\right\|_2^2 .
\label{eq:gb_opt}
\end{equation}
The problem is solved with MATLAB's \texttt{lsqnonlin} using the
trust-region-reflective algorithm, initialized from the nominal parameter vector, with
the bounds enforced directly during the search.

For each model structure \(M\), one iteration of the identification loop consists of:
\begin{enumerate}
\item \textbf{Full-order update:} map \(\bar{\theta}\) to \(\theta\) and update
\((A_{\mathrm{Full}}(\theta),B_{\mathrm{Full}}(\theta),\) \(C_{\mathrm{Full}}(\theta),D_{\mathrm{Full}}(\theta))\).
\item \textbf{Model update:} if \(M=\mathrm{Full}\), use the updated full-order
realization directly. For SP-PF and SP-Energy, recompute the slow/fast partition from
the updated realization and construct the reduced model following
Algorithms~\ref{alg:sp_pf} and \ref{alg:sp_energy}. For BT, apply
Algorithm~\ref{alg:bt_hsv}.
\item \textbf{Simulation and evaluation:} if the stability condition is satisfied,
simulate the realization and evaluate the cost in \eqref{eq:gb_opt}.
\end{enumerate}

\subsection{Performance Metrics and Comparison Criteria}
\label{subsec:performance_metrics}

Both time-domain and eigenvalue-based criteria are used to compare the identified
models: the goal is to quantify how each model reproduces the measured response and
how well it preserves the dominant small-signal dynamics. The same metrics apply to
all four structures.

\subsubsection{Time-Domain Fit}
\label{subsubsec:metric_fit}
Time-domain agreement is quantified through the output-error cost in
\eqref{eq:gb_opt}. After convergence, the identified model quality is assessed
using the normalized fit metric
\begin{equation}
\mathrm{FIT}_M = 100\left(1-\frac{\|y_{\mathrm{meas}}-y_M\|_2}{\|y_{\mathrm{meas}}-\overline{y}_{\mathrm{meas}}\|_2}\right),
\label{eq:fit_metric}
\end{equation}
where \(\overline{y}_{\mathrm{meas}}\) is the sample mean of the measured output.
Higher values of \(\mathrm{FIT}_M\) indicate better agreement between measured and
simulated transient responses. This metric, however, only evaluates time-domain
accuracy and does not guarantee preservation of the modal properties relevant to
stability analysis.

\subsubsection{Eigenvalue-Based Metrics}
\label{subsubsec:metric_modal}
Small-signal behavior is assessed from the dominant eigenvalues and their damping
ratio and oscillation frequency, defined in \eqref{eq:rom_fk_zetak}. The comparison
focuses on how well each realization preserves eigenvalue location, damping ratio,
and frequency relative to the identified FOM. Parameter sensitivity is also evaluated
by examining the displacement of the dominant eigenvalues when the identified
parameters are varied within the bounds given in Section~\ref{subsec:gb_problem}.

\subsubsection{Comparison Criteria}
\label{subsubsec:metric_comparison}
Models are compared across time-domain fit, eigenvalue preservation, parameter
sensitivity, and physical interpretability.

\section{Nominal Full-Order Model and Converter Architecture}
\label{sec:gfm_full}

\begin{figure*}[t!]
    \centering
    \vspace{-1cm}
    \includegraphics[trim={0 0 0 0},clip,width=0.925\textwidth]{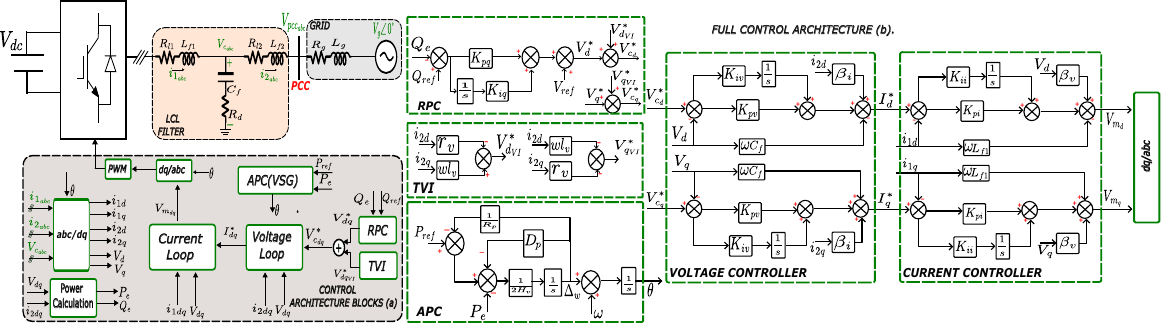}
    \caption{(a) GFM control architecture blocks (subsystems); (b) Full control architecture of the GFM converter.}
    \label{fig:gfm1}
    \label{fig:gfm2}
\end{figure*}

This section presents the control architecture of the GFM converter and derives its FOM. Fig.~\ref{fig:gfm2}(b) presents the detailed control architecture of the converter as explained in Section~\ref{subsec:control_arch}. In this structure, the active power controller is the Virtual Synchronous Generator (VSG) approach, the reactive power controller (RPC, including the transient virtual impedance (TVI)), inner voltage and current controllers (VC--CC), digital delay, LCL filter, and grid-side RL Thevenin equivalent are first modeled as individual subsystems and then interconnected through the CCM \cite{11009187}.

\begin{table}[h]
      \vspace{-0.25cm}
\caption{Design and control parameters of the GFM converter.}
\label{tab:params_design_control}
\centering
\scriptsize
\renewcommand{\arraystretch}{0.82}
\setlength{\tabcolsep}{1pt}
\begin{tabular}{@{}lll@{\hspace{1pt}}lll@{}}
\toprule
\textbf{Symbol} & \textbf{Name} & \textbf{Value} & \textbf{Symbol} & \textbf{Name} & \textbf{Value} \\
\midrule
$A_{\mathrm{nom}}$ & Apparent power & 2 kVA & $K_{pi}$ & Curr. loop P gain & 1.0161 \\
$V_{ac}$ & AC voltage & 420 V & $K_{ii}$ & Curr. loop I gain & 0.3125 \\
$V_{dc}$ & DC-link voltage & 750 V & $\beta_V$ & Voltage feedfwd. & 1 \\
$\omega_n$ & Nominal freq. & 314.16 rad/s & $\beta_I$ & Current feedfwd. & 1 \\
$f_{sw}$ & Switching freq. & $20\times10^3$ Hz & $r_v$ & Virtual resist. & 0.0025 $\Omega$ \\
$P_{\mathrm{ref}}$ & Active power ref. & 1 p.u. & $l_v$ & Virtual induct. & $5\times10^{-6}$ H \\
$Q_{\mathrm{ref}}$ & Reactive power ref. & 0.1 p.u. & $R_{f1}$ & Filter resist. 1 & 0.0995 $\Omega$ \\
$H_v$ & Virtual inertia & 0.15 s & $L_{f1}$ & Filter induct. 1 & 0.001 H \\
$R_p$ & Active power droop & 0.015 & $R_{f2}$ & Filter resist. 2 & 0 $\Omega$ \\
$D_p$ & Damping coeff. & 0.66 & $L_{f2}$ & Filter induct. 2 & $3.48\times10^{-5}$ H \\
$K_{pq}$ & React. droop gain & 0.042 & $C_f$ & Filter capac. & $1.5\times10^{-5}$ F \\
$K_{iq}$ & React. integral gain & $1\times10^{-4}$ & $R_d$ & Damping resist. & 0.746 $\Omega$ \\
$K_{pv}$ & Voltage loop P gain & $2.12\times10^{-5}$ & $L_g$ & Grid inductance & 0.004 H \\
$K_{iv}$ & Voltage loop I gain & 120 & $R_g$ & Grid resistance & 2.05 $\Omega$ \\
\bottomrule
\end{tabular}
\end{table}

\subsection{Control Architecture}
\label{subsec:control_arch}

The converter is modeled as an interconnection of the subsystems described below,
each expressed in small-signal form around the operating point. The measured active
and reactive powers \((P_e,Q_e)\) are computed in the IRF from the capacitor node
voltage and the grid-side current. Throughout the subsystem realizations that follow, \(J=\begin{bmatrix}0&-1\\1&0\end{bmatrix}\).

The synchronization dynamics procedure are governed by the VSG approach with virtual inertia \(H_v\), damping \(D_p\), and active power droop \(R_p\) together with \(\dot{\delta}=\omega-\omega_n\), where \(\delta\) is the angle between
the IRF and the GRF.
The linearized VSG realization used in the CCM assembly is
\begin{equation}
\begin{aligned}
\Delta\dot{\omega} &= -\left(\frac{1}{2H_v\omega_nR_p}+\frac{D_p}{2H_v}\right)\Delta\omega +\frac{1}{2H_v\omega_n}\Delta P_{\mathrm{ref}} \\
&\quad -\frac{1}{2H_v\omega_n}\Delta P_e,\\
\Delta\dot{\delta} &=\Delta\omega.
\end{aligned}
\label{eq:app_vsg}
\end{equation}

Reactive power regulation generates a voltage magnitude reference via a droop plus
integral characteristic with gains \((K_{pq},K_{iq})\). A TVI with parameters \((r_v,l_v)\) modifies
this reference using the grid-side current. The TVI is implemented as algebraic formula within the RPC block and therefore adds no states. The linearized RPC+TVI realization is
\begin{align}
\Delta \dot{\phi}_q &= \Delta Q_{\mathrm{ref}}-\Delta Q_e, \label{eq:app_rpc_int}\\
\Delta E^\ast &= K_{iq}\Delta \phi_q + K_{pq}(\Delta Q_{\mathrm{ref}}-\Delta Q_e), \label{eq:app_rpc_E}\\
\Delta \mathbf{v}^{\ast c}_{c,dq}(s) &= \begin{bmatrix} \Delta E^\ast(s)\\ 0 \end{bmatrix} - \big((r_v+s l_v)I+\omega_n l_v J\big)\Delta\mathbf{i}^c_{2,dq}(s). \label{eq:app_rpc_out}
\end{align}

The inner control employs cascaded PI voltage and current controllers with feedforward and couplings, with gains \((K_{pv},K_{iv})\) and \((K_{pi},K_{ii})\) and
feedforward terms \(\beta_I\) and \(\beta_V\). The linearized VC realization is
\begin{align}
\Delta \dot{\boldsymbol{\phi}}_v &= \Delta \mathbf{v}^{\ast c}_{c,dq} - \Delta \mathbf{v}^c_{c,dq}, \label{eq:app_vc_int}\\
\Delta \mathbf{i}^{\ast c}_{1,dq} &= K_{iv}\Delta \boldsymbol{\phi}_v + K_{pv}\!\left(\Delta \mathbf{v}^{\ast c}_{c,dq}-\Delta \mathbf{v}^c_{c,dq}\right) \nonumber\\
&\quad -\omega_n C_f J\,\Delta \mathbf{v}^c_{c,dq} +\beta_I\Delta \mathbf{i}^c_{2,dq}. \label{eq:app_vc_out}
\end{align}
and the linearized CC realization is
\begin{align}
\Delta \dot{\boldsymbol{\phi}}_i &= \Delta \mathbf{i}^{\ast c}_{1,dq} - \Delta \mathbf{i}^c_{1,dq}, \label{eq:app_cc_int}\\
\Delta \mathbf{v}_{m,dq} &= K_{ii}\Delta \boldsymbol{\phi}_i + K_{pi}\!\left(\Delta \mathbf{i}^{\ast c}_{1,dq}-\Delta \mathbf{i}^c_{1,dq}\right) \nonumber\\
&\quad -\omega_n L_{f1} J\,\Delta \mathbf{i}^c_{1,dq} +\beta_V\Delta \mathbf{v}^c_{c,dq}. \label{eq:app_cc_out}
\end{align}

The delay between the modulating command \(\Delta \mathbf{v}_{m,dq}\) and the inverter terminal voltage \(\Delta \mathbf{v}^{c}_{dq,\mathrm{inv}}\) is implemented using a third-order Padé approximation, as in \cite{11009187}, applied to both \(dq\)
channels and contributing the six states \(x_{\mathrm{del}1}\) to
\(x_{\mathrm{del}6}\).

The LCL filter and the grid network are expressed in the GRF, whereas the control subsystems are expressed in the IRF. Their coupling is established through the linearized reference-frame interface. The grid RL Thevenin equivalent is absorbed into the grid-side branch of the LCL filter, so that
\(L_{t2}=L_{f2}+L_g\) and \(R_{t2}=R_{f2}+R_g\). This subsystem contributes the
converter-side current \(\mathbf{i}^{s}_{1,dq}\), the capacitor branch voltage
\(\mathbf{v}^{s}_{cf,dq}\), and the grid-side current \(\mathbf{i}^{s}_{2,dq}\).
The voltage at PCC remains as a required output variable and is recovered as an
algebraic output.

The linearized subsystems are stacked into block-diagonal matrices \cite{11009187}. The input vector is
\begin{equation}
\Delta u_{sys}
=
\begin{bmatrix}
\Delta P_{\mathrm{ref}} &
\Delta Q_{\mathrm{ref}} &
\Delta v^s_{cd} &
\Delta v^s_{cq}
\end{bmatrix}^{\top}.
\label{eq:sec3_usys}
\end{equation}

\begin{figure}[t]
      \vspace{-0.25cm}
    \centering
    \includegraphics[trim={0 7 15 5},clip,width=0.85\columnwidth]{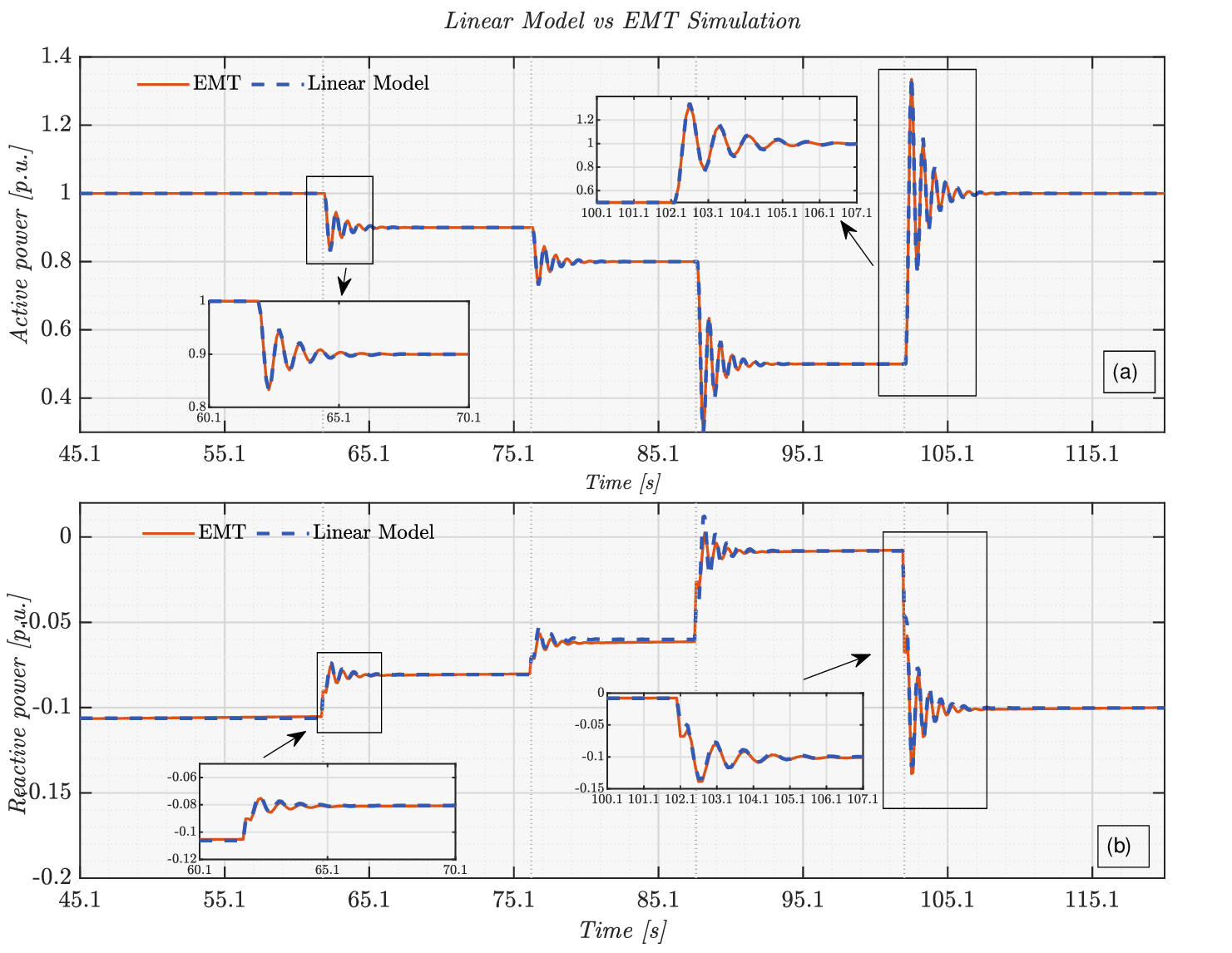}
    \caption{Comparison of the active (a) and reactive (b) power responses from the linearized CCM-based model and the nonlinear EMT simulation.}
    \label{fig:lmvssim}
\end{figure}

All signal level interconnections among subsystem ports (measurement maps, TVI reference modification, delay actuation, LCL--grid coupling, and linearized IRF--GRF reference-frame mappings) are collected into the affine CCM relations \cite{11009187}. The unified full-order realization is \eqref{eq:prob_fom}, which collects the output channels of interest selected for identification and validation.
The full-order linearized model used throughout the paper contains 19 states: \(\omega\), \(\delta\), \(\phi_q\), \(\phi_{vd}\), \(\phi_{vq}\), \(\phi_{id}\), \(\phi_{iq}\), \(x_{\mathrm{del}1}\), \(x_{\mathrm{del}2}\), \(x_{\mathrm{del}3}\), \(x_{\mathrm{del}4}\), \(x_{\mathrm{del}5}\), \(x_{\mathrm{del}6}\), \(i_{1d}\), \(i_{1q}\), \(v_{c,d}\), \(v_{c,q}\), \(i_{2d}\), and \(i_{2q}\).
The control architecture shown in Fig.~\ref{fig:gfm2}(b) was implemented in Simulink using the nominal parameters reported in Table~\ref{tab:params_design_control}.
The derived small-signal model was compared with the nonlinear EMT simulation, and the match shown in Fig.~\ref{fig:lmvssim} supports the use of this realization.
\section{Experimental Setup and Data Acquisition }
\label{sec:exp}

\begin{figure}[h]
      \vspace{-1cm}
    \centering
    \begin{minipage}[c]{0.45\columnwidth}
        \centering
        \includegraphics[trim={0 5 0 5},clip,angle=-90,width=\linewidth]{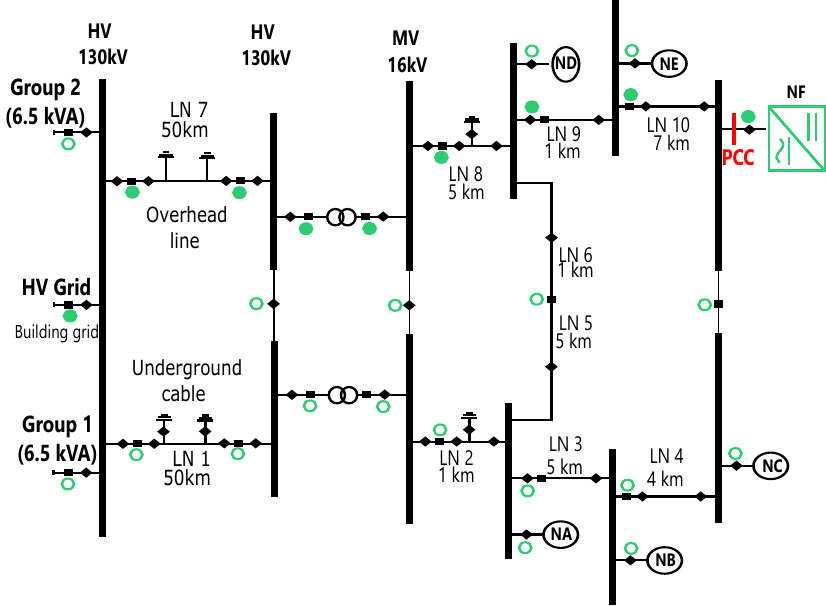}\\
        (a)
    \end{minipage}
    \hspace{0.01\columnwidth}
    \begin{minipage}[c]{0.45\columnwidth}
        \centering
        \includegraphics[trim={0 0 0 50},clip,width=\linewidth]{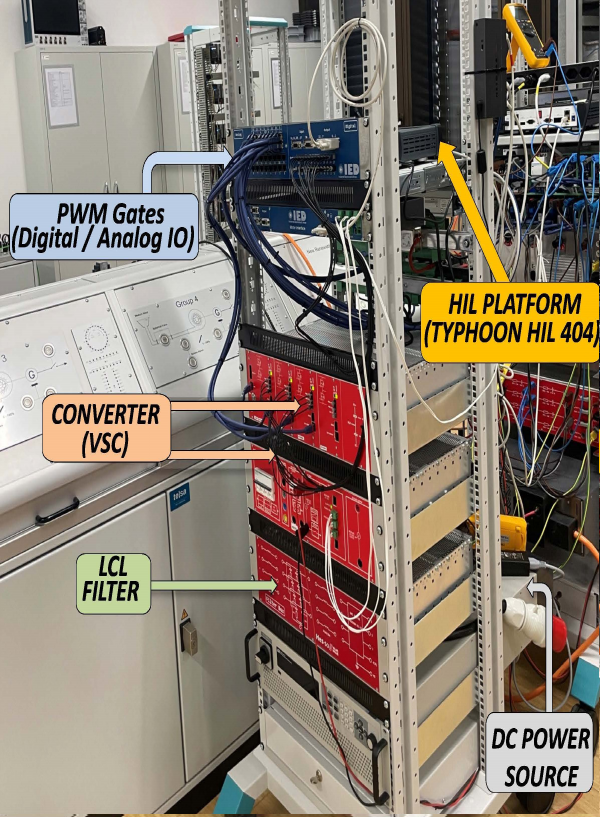}\\
        (b)
    \end{minipage}
    \caption{(a) Dispatch topology testbench. (b) Experimental setup.}
    \label{fig:dispatch_topology}
    \label{fig:power_setup}
\end{figure}

The experimental validation is carried out at the Gridlab laboratory of HES-SO Valais-Wallis. The adopted facility reproduces a reduced-scale medium-voltage transmission grid with topology shown in Fig.~\ref{fig:dispatch_topology}(a).

\begin{figure*}[t!]
    \centering
    \vspace{-1.2cm}
    \includegraphics[trim={0 8 0 39},clip,width=0.925\textwidth]{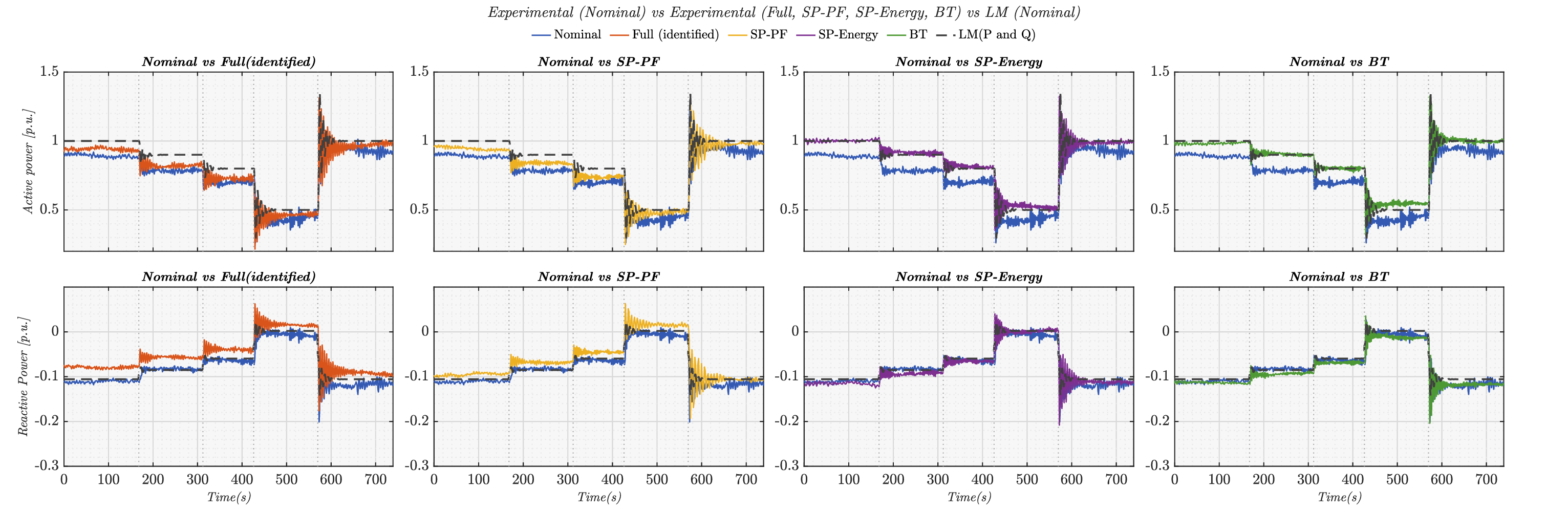}
    \caption{Time-domain comparison of measured and identified active and reactive power responses for the full-order and reduced-order models.}
    \label{fig:time_domain_validation}
\end{figure*}

In the configuration considered here, the converter is connected at node F, thus the analysis is limited to the network section formed by Lines 7 to 10, whose aggregated electrical parameters correspond to the grid resistance $R_g$ and grid inductance $L_g$ reported in Table~\ref{tab:params_design_control}. The system operates as a three phase line voltage of $420~\mathrm{V}_{\mathrm{RMS}}$ and, although the testbench also supports islanded operation, all the experiments reported in this work were performed in grid-connected mode, where the building grid acts as a stiff reference for the GFM converter.
The power interface is a three phase converter. It includes LCL filters, EMI filters, and DC-link precharge circuitry. The control platform supports either a TI-TMS320F28377S microprocessor based controller or a Typhoon Hardware-in-the-Loop system; in the present setup, the converter is controlled through the Typhoon HIL. The DC link is supplied by a regulated $750~\mathrm{V}_{\mathrm{DC}}$ source.
The control runs in real time and is interfaced with the physical power stage, forming the setup shown in Fig.~\ref{fig:power_setup}(b). RMS voltage and current measurements are acquired through SICAM-T devices. The implementation is validated through the active and reactive power responses shown in Fig.~\ref{fig:time_domain_validation} (Nominal, blue curve), which illustrate the behavior of the grid-connected GFM converter under $P_{ref}$ disturbances, with \(P_{\mathrm{ref}}\) varied from 1 to 0.9, 0.8, 0.5, and back to 1 p.u. This sequence provides multiple disturbances of different amplitudes within the same experiment. The resulting sampled dataset is used for validation and identification in Section~\ref{sec:greybox}, and for the linearized and nonlinear simulations discussed in Section~\ref{sec:gfm_full}.

\section{Results and Discussion}
\label{sec:results}

This section presents the experimental identification results obtained from the grid-connected GFM converter and compares Full with the ROMs.

\begin{figure*}[t!]
    \centering
      \vspace{-0.5cm}
    \includegraphics[trim={0 10 0 8},clip,width=0.925\textwidth]{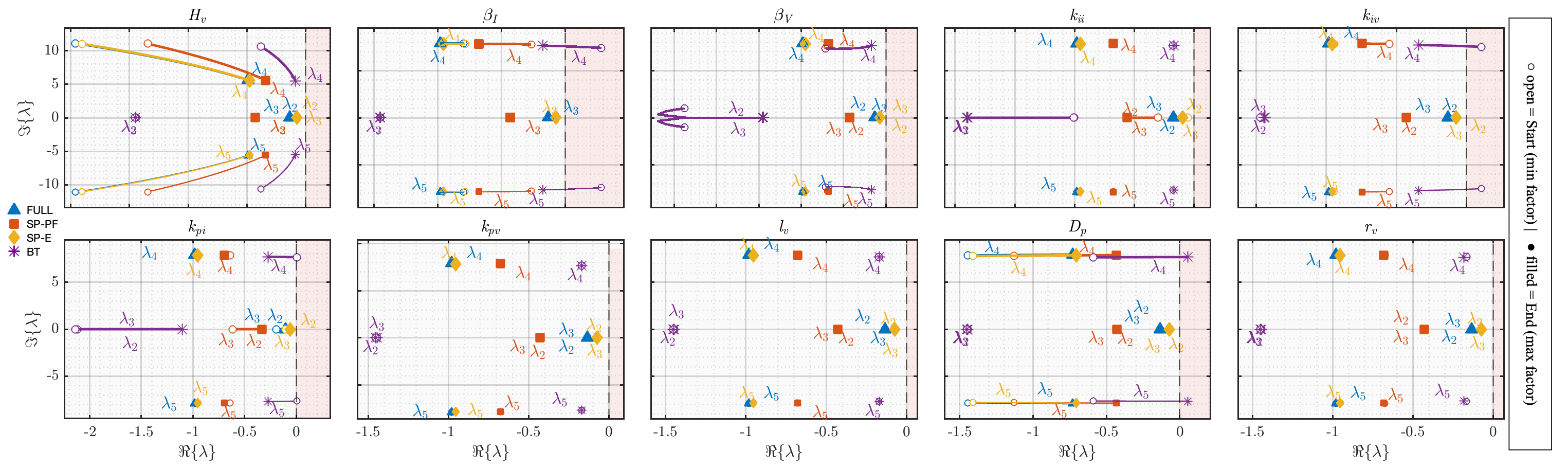}
    \caption{Pole displacement of the full-order and reduced-order models obtained by sweeping each selected degree of freedom across its admissible identification range; open markers denote the lower bound, filled markers the upper bound.}
    \label{fig:eigenvalues_map}
\end{figure*}

\subsection{Reduced Model Order and Retained States}

\begin{figure}[h]
      \vspace{-0.6cm}
    \centering
    \includegraphics[trim={40 52 15 28},clip,width=0.85\columnwidth]{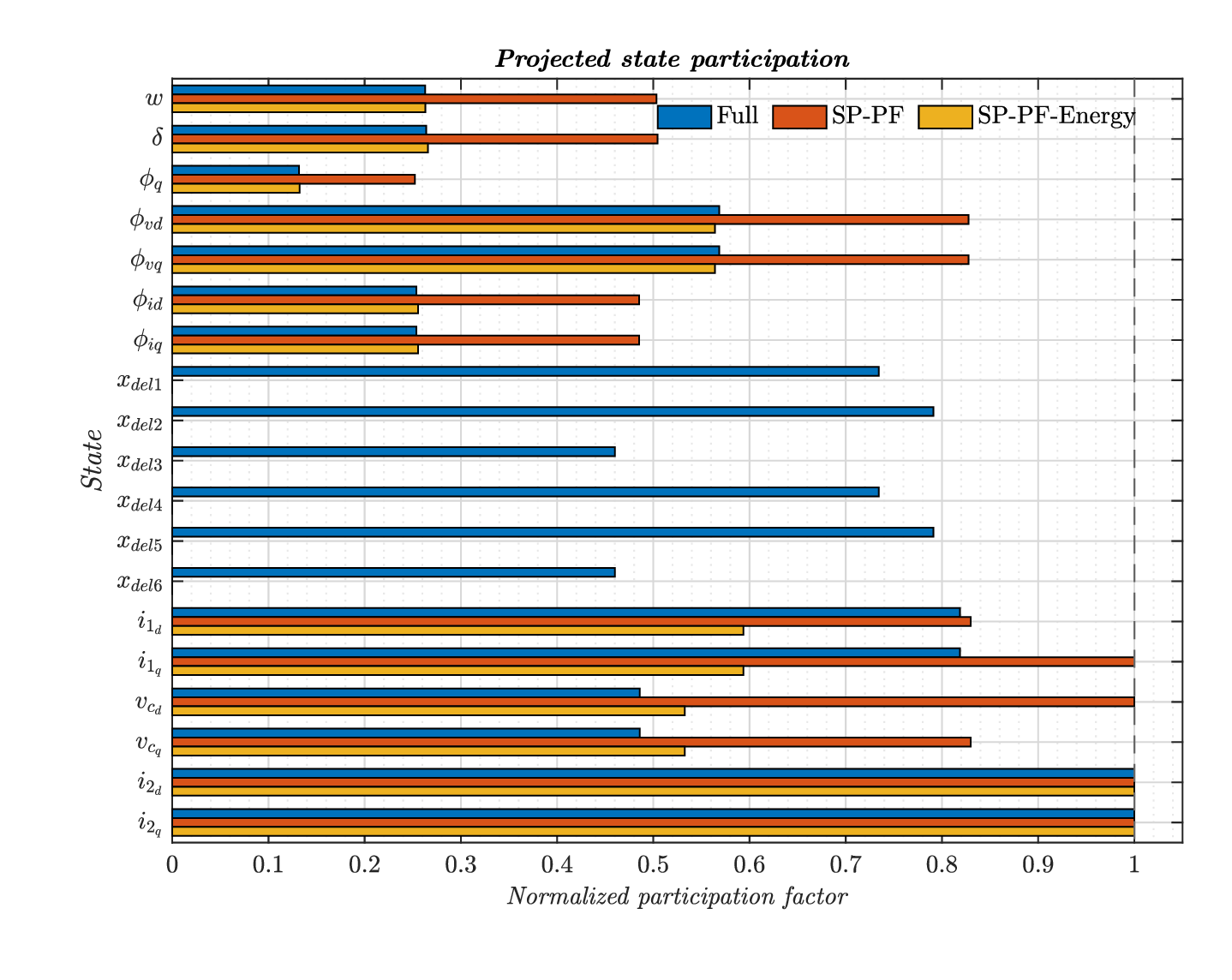}
    \caption{Projected participation factors of the full-order and SP-based ROMs.}
    \label{fig:PF}
\end{figure}

Fig.~\ref{fig:PF} shows the projected normalized participation factors of the SP-based reduced models in the original state coordinates. Both the SP-PF and SP-Energy model retain 13 physical states by removing the six delay states mentioned in Section \ref{subsec:control_arch}. When projected back onto the original coordinates, the eliminated states appear with zero participation. Both criteria therefore eliminate the same six states, but they assign different
weights to the retained ones. This difference arises because the slow/fast partition is recomputed at every identification iteration (Section~\ref{sec:greybox}) from the fixed frequency cutoff and modal energy criteria, respectively; as $\theta$ changes during the search, the two criteria can classify states differently, which explains the different parameter estimates and modal properties in Table~\ref{tab:params_models_v3_test} and Table~\ref{tab:eigs_freq_zeta_fullcomp}. BT does not appear in Fig.~\ref{fig:PF} because its reduction is carried out in balanced coordinates rather than by retaining a subset of physical states. The BT model is therefore characterized only by its retained order \(r\). The selected BT order is \(r=10\), corresponding to the smallest order that satisfies the prescribed cumulative Hankel singular value threshold.

\begin{table}[t!]
      \vspace{-1cm}
\centering
\scriptsize
\caption{Comparison of the three dominant states associated with the critical modes of the nominal and SP-based ROMs.}
\label{tab:dominant_states_critical_modes}
\renewcommand{\arraystretch}{0.78}
\setlength{\tabcolsep}{3pt}
\begin{tabular}{c c c c}
\toprule
\textbf{Mode} & \textbf{Nominal} & \textbf{SP-PF} & \textbf{SP-Energy} \\
\midrule
$\lambda_{2}$ 
& $\phi_{id},\; \phi_{iq},\; \phi_{vd}$
& $\phi_{id},\; \phi_{iq},\; \phi_{vd}$
& $\phi_{id},\; \phi_{iq},\; \phi_{vd}$ \\

$\lambda_{3}$ 
& $\phi_{iq},\; \phi_{id},\; \phi_{vq}$
& $\phi_{iq},\; \phi_{id},\; \phi_{vq}$
& $\phi_{iq},\; \phi_{id},\; \phi_{vq}$ \\

$\lambda_{4}$ 
& $\delta,\; w,\; i_{2_q}$
& $\delta,\; w,\; \phi_{vd}$
& $\delta,\; w,\; i_{2_q}$ \\

$\lambda_{5}$ 
& $\delta,\; w,\; i_{2_q}$
& $\delta,\; w,\; \phi_{vd}$
& $\delta,\; w,\; i_{2_q}$ \\

$\lambda_{6}$ 
& $\phi_{vq},\; \phi_{vd},\; i_{2_d}$
& $\phi_{vq},\; \phi_{vd},\; v_{c_q}$
& $i_{1_q},\; i_{1_d},\; \phi_{vq}$ \\

$\lambda_{7}$ 
& $\phi_{vq},\; \phi_{vd},\; i_{2_d}$
& $\phi_{vq},\; \phi_{vd},\; v_{c_q}$
& $i_{1_q},\; i_{1_d},\; \phi_{vq}$ \\

\bottomrule
\end{tabular}
\end{table}

\subsection{Parameter Estimation Results}

\begin{table}[t]
\caption{Identified parameter values for the nominal model, Full, and the reduced-order realizations.}
\label{tab:params_models_v3_test}
\centering
\scriptsize
\renewcommand{\arraystretch}{0.82}
\setlength{\tabcolsep}{1.5pt}
\begin{tabular}{lccccc}
\toprule
& \textbf{Nominal} & \textbf{Full (identified)} & \textbf{SP-PF} & \textbf{SP-Energy} & \textbf{BT} \\
\midrule
$H_v$ (s) & 0.150 & 0.152 & 0.155 & 0.155 & 0.16 \\
$R_p$ & 0.015 & 0.015 & 0.015 & 0.015 & 0.015 \\
$\beta_I$ & 1.00 & 0.99 & 0.66 & 0.90 & 0.63 \\
$\beta_V$ & 1.00 & 0.7 & 0.87 & 1 & 0.92 \\
$K_{ii}$ & 0.31 & 0.35 & 0.33 & 0.32 & 0.28 \\
$K_{iq}$ & $1\times10^{-4}$ & $1\times10^{-4}$ & $1\times10^{-4}$ & $1\times10^{-4}$ & $1\times10^{-4}$ \\
$K_{iv}$ & 120.0 & 119.41 & 34.63 & 20.24 & 23.92 \\
$K_{pi}$ & 1.02 & 0.3898 & 0.77 & 4.45 & 0.198 \\
$K_{pq}$ & 0.042 & 0.042 & 0.042 & 0.042 & 0.042 \\
$K_{pv}$ & $2.1\times10^{-5}$ & $7.02\times10^{-5}$ & $5.58\times10^{-5}$ & $1.1\times10^{-6}$ & $5.7\times10^{-5}$ \\
$l_v$ & $5\times10^{-6}$ & $5\times10^{-6}$ & $5\times10^{-6}$ & $5\times10^{-6}$ & $5\times10^{-6}$ \\
$D_p$ & 0.66 & 0.66 & 0.66 & 0.66 & 0.66 \\
$r_v$ & 0.0025 & 0.0025 & 0.0034 & 0.0031 & 0.0029 \\
\bottomrule
\end{tabular}
\end{table}
Fig.~\ref{fig:time_domain_validation} compares the measured data collected 
response with the responses of the identified FOM and ROM. All models were identified from the same input/output dataset, which 
allows a comparison of the estimated parameter values across the model 
structures. The identified parameter values are summarized in 
Table~\ref{tab:params_models_v3_test}. Comparing the nominal model with Full reveals 
how the estimated parameters deviate from their nominal initial values 
reported in Table~\ref{tab:params_design_control}. The feedforward gains 
\((\beta_I,\beta_V)\) are identified as \(0.99\) and \(0.70\) with Full, instead of their nominal unity values. For the reduced-order 
models, \(\beta_I\) is identified as \(0.66\) in SP-PF, \(0.90\) in 
SP-Energy, and \(0.63\) in BT, while \(\beta_V\) is identified as \(0.87\) 
in SP-PF, \(1.00\) in SP-Energy, and \(0.92\) in BT. The transient virtual 
impedance parameters remain close to their nominal values across all 
realizations. In Full, \(l_v\) and \(r_v\) are 
identified as \(5\times10^{-6}\) and \(0.0025\), consistent with the base 
model values. Some parameters take the same values in Full and ROMs. In particular, the active power droop \(R_p=0.015\), the 
reactive integral gain \(K_{iq}=1\times10^{-4}\), and the reactive droop 
\(K_{pq}=0.042\) are identical in Full and in the ROMs. Additionally, the damping coefficient \(D_p=0.66\) is 
consistent across all model structures, matching the nominal value.

On the other hand, the largest differences across model structures appear 
in the virtual inertia \(H_v\) and in the voltage loop gains. Full gives 
\(H_v=0.152~\mathrm{s}\), whereas the ROMs identify 
\(0.155~\mathrm{s}\) for SP-PF, \(0.155~\mathrm{s}\) for SP-Energy, and 
\(0.160~\mathrm{s}\) for BT. Among all models, the inertia estimates remain 
close to the nominal value of \(0.150~\mathrm{s}\). A similar behavior 
appears in \(K_{iv}\), for which Full gives 
\(119.41\), while SP-PF, SP-Energy, and BT give \(34.63\), \(20.24\), and 
\(23.92\), respectively. For \(K_{pi}\), Full gives \(0.390\), compared 
with \(0.77\) in SP-PF, \(4.45\) in SP-Energy, and \(0.198\) in BT. This 
is also observed in \(K_{pv}\), which remains of order \(10^{-5}\) in Full, SP-PF, and BT, but changes to values of order 
\(10^{-6}\) in SP-Energy.
In general, differences between model structures are centered in the 
feedforward gains and in the voltage loop gains. In this comparison, 
\(H_v\) and \(D_p\) are robustly identified across all structures, while 
\(K_{iv}\), \(K_{pi}\), and \(K_{pv}\) show the largest variation.

\begin{table*}[t!]
\vspace{-1cm}
\caption{Comparison of eigenvalues for the nominal model, Full, and ROMs, and the natural frequency $f_n$ [Hz] and damping ratio $\zeta$.}
\label{tab:eigs_freq_zeta_fullcomp}
\centering
\renewcommand{\arraystretch}{0.82}
\setlength{\tabcolsep}{1.2pt}
\scriptsize

\begin{tabular}{lccc ccc ccc ccc ccc}
\toprule
\textbf{$\lambda$} &
\multicolumn{3}{c}{\textbf{Nominal}} &
\multicolumn{3}{c}{\textbf{Full (identified)}} &
\multicolumn{3}{c}{\textbf{SP-PF}} &
\multicolumn{3}{c}{\textbf{SP-Energy}} &
\multicolumn{3}{c}{\textbf{BT}} \\
\cmidrule(lr){2-4}\cmidrule(lr){5-7}\cmidrule(lr){8-10}
\cmidrule(lr){11-13}\cmidrule(lr){14-16}
& $\lambda$ & $f_n$ & $\zeta$
& $\lambda$ & $f_n$ & $\zeta$
& $\lambda$ & $f_n$ & $\zeta$
& $\lambda$ & $f_n$ & $\zeta$
& $\lambda$ & $f_n$ & $\zeta$ \\
\midrule

$\lambda_{2}$
& $-0.308$ & $0.049$ & $1.000$
& $-0.133$ & $0.021$ & $1.000$
& $-0.429$ & $0.068$ & $1.000$
& $-0.072$ & $0.011$ & $1.000$
& $-1.446$ & $0.230$ & $1.000$ \\

$\lambda_{3}$
& $-0.308$ & $0.049$ & $1.000$
& $-0.133$ & $0.021$ & $1.000$
& $-0.430$ & $0.068$ & $1.000$
& $-0.072$ & $0.011$ & $1.000$
& $-1.453$ & $0.231$ & $1.000$ \\

$\lambda_{4}$
& $-1.494{+}30.881i$ & $4.921$ & $0.048$
& $-0.981{+}7.928i$ & $1.272$ & $0.123$
& $-0.676{+}7.873i$ & $1.258$ & $0.086$
& $-0.953{+}7.861i$ & $1.260$ & $0.120$
& $-0.170{+}7.682i$ & $1.223$ & $0.022$ \\

$\lambda_{5}$
& $-1.494{-}30.881i$ & $4.921$ & $0.048$
& $-0.981{-}7.928i$ & $1.272$ & $0.123$
& $-0.676{-}7.873i$ & $1.258$ & $0.086$
& $-0.953{-}7.861i$ & $1.260$ & $0.120$
& $-0.170{-}7.682i$ & $1.223$ & $0.022$ \\

$\lambda_{6}$
& $-103.2{+}431.8i$ & $70.658$ & $0.232$
& $-292.4{+}363.0i$ & $74.185$ & $0.627$
& $-102.9{+}43.2i$  & $17.764$ & $0.922$
& $-421.9{+}372.4i$ & $89.562$ & $0.750$
& $-225.8{+}58.0i$  & $37.097$ & $0.969$ \\

$\lambda_{7}$
& $-103.2{-}431.8i$ & $70.658$ & $0.232$
& $-292.4{-}363.0i$ & $74.185$ & $0.627$
& $-102.9{-}43.2i$  & $17.764$ & $0.922$
& $-421.9{-}372.4i$ & $89.562$ & $0.750$
& $-225.8{-}58.0i$  & $37.097$ & $0.969$ \\

\bottomrule
\end{tabular}
\end{table*}
\subsection{Eigenvalue Comparison and Stability}
\begin{figure}[h]
      \vspace{-0.15cm}
    \centering
    \includegraphics[trim={20 10 50 16},clip,width=0.85\columnwidth]{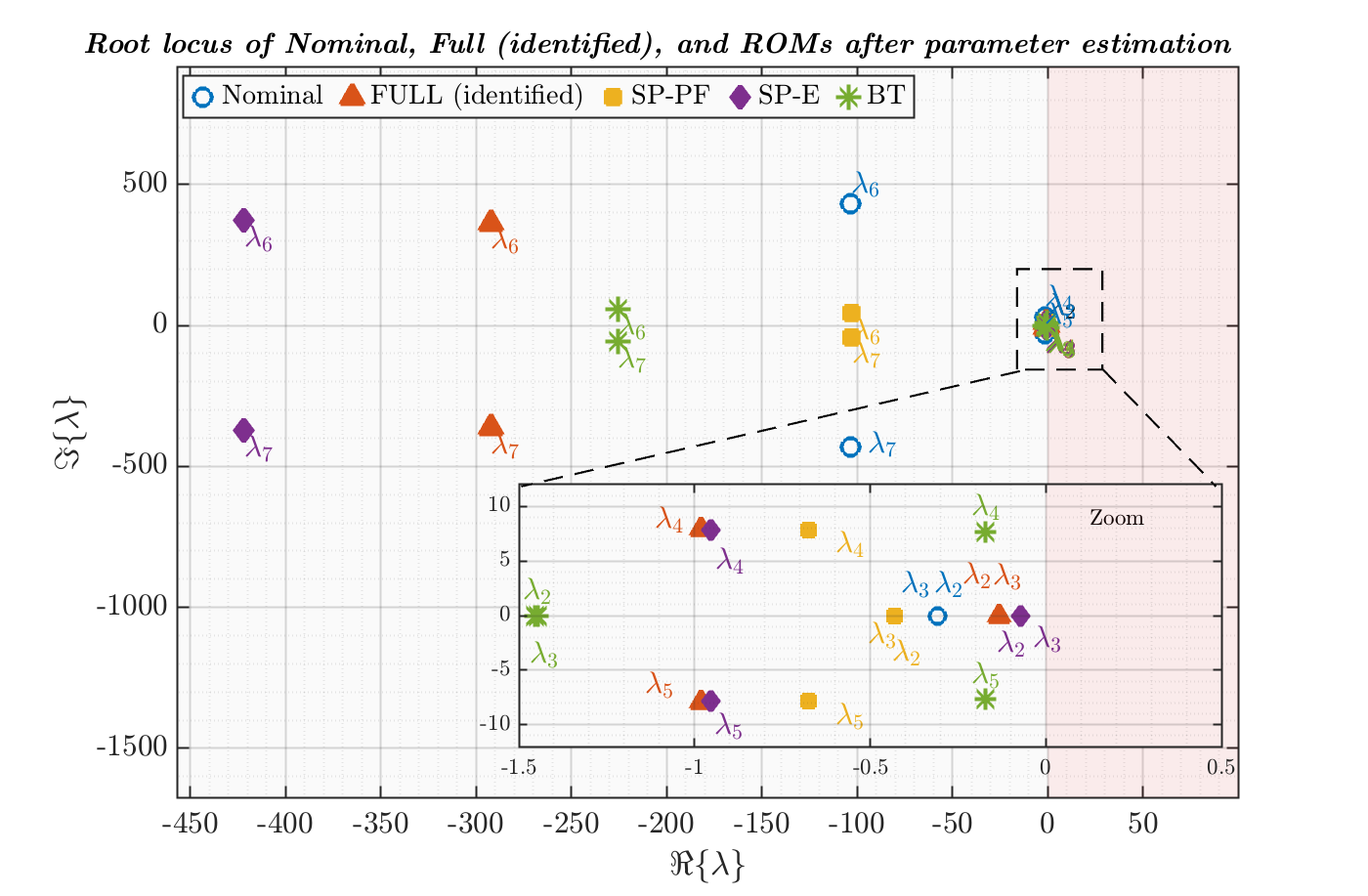}
    \caption{Root locus of Full and ROMs after parameter estimation.}
    \label{fig:root_locus}\end{figure}

The dominant modes are compared numerically in
Table~\ref{tab:eigs_freq_zeta_fullcomp} and graphically in
Fig.~\ref{fig:root_locus}. In this sense, BT can be compared directly with Full and the SP-based
models at the modal level. The first real pair, \((\lambda_2,\lambda_3)\),
shifts from about \(-0.308\) in the nominal model to \(-0.133\) in Full. In SP-Energy, this pair moves closer 
to the origin, reaching about \(-0.072\), while SP-PF moves it in the 
opposite direction to \(-0.429\), and BT places it far from the origin 
at \(-1.446\).
The pair \((\lambda_4,\lambda_5)\) appears at \(4.92~\mathrm{Hz}\) in 
the nominal model, with a damping ratio of \(\zeta=0.048\). After the 
identification framework, the Full places this mode at \(1.27~\mathrm{Hz}\) 
and increases the damping to \(\zeta=0.123\). The ROMs shift the frequency 
to the range between \(1.22\) and \(1.26~\mathrm{Hz}\). Regarding damping, 
SP-PF gives \(\zeta=0.086\), SP-Energy gives \(\zeta=0.120\), and BT gives 
\(\zeta=0.022\). Among these results, SP-Energy more accurately reproduces 
the damping of this mode, while BT moves the mode closer to the stability 
limit as highlighted in Fig.~\ref{fig:root_locus}. The higher-frequency 
pair \((\lambda_6,\lambda_7)\) also changes after identification. In the 
nominal model, it appears near \(70.7~\mathrm{Hz}\) with \(\zeta=0.232\). 
In Full, the frequency increases to about 
\(74.2~\mathrm{Hz}\) and the damping increases to \(\zeta=0.627\). In the
ROMs, this pair appears at \(17.8~\mathrm{Hz}\) in SP-PF, 
\(89.6~\mathrm{Hz}\) in SP-Energy, and \(37.1~\mathrm{Hz}\) in BT. The 
damping ratios remain close to the value of Full, but the frequencies 
differ noticeably across the model structures.

\subsection{Pole Sensitivity Within the Identification Bounds}
Fig.~\ref{fig:eigenvalues_map} illustrates how the dominant eigenvalues move when each selected parameter is swept over its selected range as given in Section~\ref{subsec:gb_problem}, while the remaining parameters are kept fixed. For each parameter, the open marker indicates the pole location at the lower bound and the filled marker indicates the location at the upper bound.

As seen in the plots, sweeping parameters like $K_{iv}$ and $H_v$ causes much wider pole displacements in the BT model. A clear example is \(\beta_I\), whose effect differs markedly among the four structures. In the present case, sweeping \(\beta_I\) over its admissible range leads to unstable realizations in Full and BT, while the SP-based models remain stable over the same interval (indicated by the red shaded region in Fig.~\ref{fig:eigenvalues_map}). For other parameters, the main difference lies in the size of the pole displacement. The parameter \(H_v\) shows one of the widest variations in pole location. For \(K_{iv}\), \(D_p\), and \(K_{pi}\), the largest motion is concentrated in the pair \((\lambda_4,\lambda_5)\). In these cases, Full and BT generally exhibit wider displacements than SP-PF and SP-Energy. Notably, sweeping \(K_{iv}\), \(K_{pi}\), and \(D_p\) in the BT realization drives the pair \((\lambda_4,\lambda_5)\) toward the unstable region, whereas the SP-based models remain within a stable region over the same interval.

\subsection{Overall Comparison of the Reduced-Order Models}
In the present comparison, the highest fit after framework is obtained by BT,
with \(FIT=75.23\%\), followed by SP-Energy with \(FIT=70.57\%\), SP-PF with 
\(FIT=68.41\%\), and Full with \(FIT=58.25\%\). 
The nominal model gives the lowest fit, at \(FIT=46.25\%\). 
These results show that all identified models improve the measured-response 
match with respect to the nominal model. In particular, BT achieves the highest 
fit, while the SP-based models give higher values than Full. This is consistent with the redistribution of state participation shown in Fig.~\ref{fig:PF}: each reduction yields a realization in which the retained states carry different weights than in the FOM, and the estimation is performed on that realization.

\section{Conclusion}
\label{sec:conclusion}
In this paper, we proposed an experimentally validated grey-box
identification framework for GFM converters, combining
a CCM-based full-order small-signal model with reduced-order realizations
obtained through SP-PF, the proposed SP-Energy method, and balanced
truncation. In time-domain terms, BT achieves the highest fit to the measured experimental
response, followed by SP-Energy and SP-PF, while Full
provides a lower fit than the ROMs; however, the modal analysis
shows that this does not correspond to the best preservation
of the small-signal dynamics. BT shifts the dominant
oscillatory mode closer to the stability boundary, with
the smallest damping margin and the highest sensitivity to parameter variations among the ROMs. The proposed SP-Energy realization provides a more balanced result, with a better fit than SP-PF and an estimated inertia and damping closer to Full.

Parameter estimates are tied to the input-output richness of the experiment: with one excited input, the active power reference, and two measured outputs, some estimates are best interpreted as coupled combinations rather than individually recovered values. This reflects the experimental data, not the framework; the same workflow extends directly to additional excited inputs, outputs, and parameter estimators, sharpening individual identifiability without changing the methodology. The SP-Energy threshold and frequency range are likewise fixed for this case study; sweeping them, and comparing sensitivity to estimation uncertainty rather than to design bounds, are direct extensions of the same framework.

\ifCLASSOPTIONcaptionsoff
  \newpage
\fi

\bibliographystyle{ieeetr}
\bibliography{biblio.bib}

\begin{IEEEbiography}[{\includegraphics[width=1in,height=1.25in,clip,keepaspectratio]{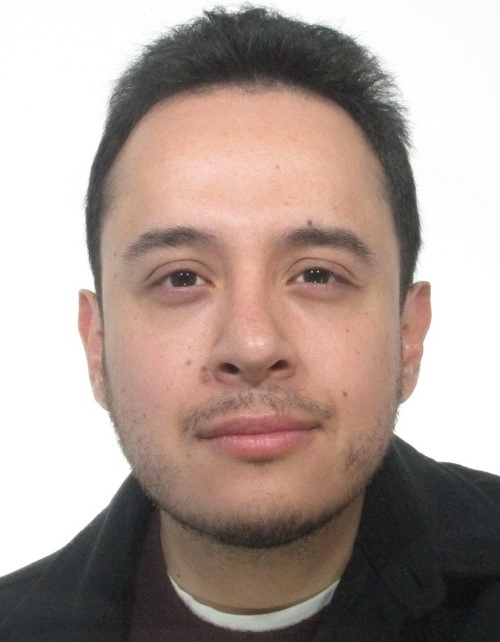}}]{Marcelo Cardenas Calle}
(Student Member, IEEE) is a Ph.D. candidate with the University of Genoa, Genoa, Italy. He holds a Master of Science in Electrical Engineering from the Budapest University of Technology and Economics, Budapest, Hungary, an MBA from Formato Educativo Business School, and a Bachelor of Engineering in Electrical Engineering from the University of Cuenca, Cuenca, Ecuador. During 2022, he worked at Centrosur, an Ecuadorian distribution system operator (DSO). He is a member of the Swiss Next Generation Network (NGN) for CIGRE. His research interests include the integration of renewable energy resources and the stability of modern electrical power grids.
\end{IEEEbiography}
\begin{IEEEbiography}[{\includegraphics[width=1in,height=1.25in,clip,keepaspectratio]{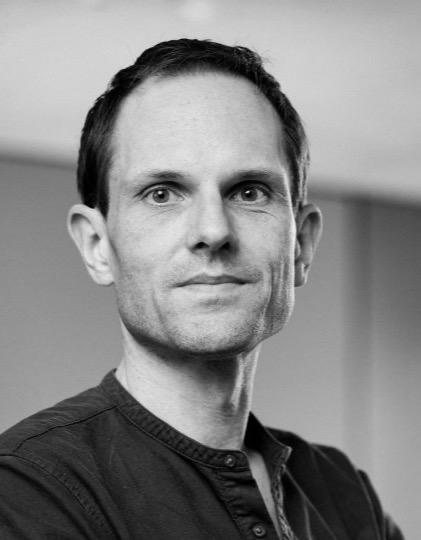}}]{Stephan Kenzelmann}
received the M.S. and Ph.D. degrees from the École Polytechnique Fédérale de Lausanne (EPFL), Switzerland, in 2007 and 2012, respectively. From 2013 to 2021, he worked in the traction development department at ABB, most recently as Principal Engineer, focusing on power electronics control. Since 2021, he has been an Associate Professor at HES-SO Valais-Wallis in Sion, Switzerland, where he is responsible for the Industrial Electronics and Drives Laboratory.
\end{IEEEbiography}
\begin{IEEEbiography}[{\includegraphics[width=1in,height=1.25in,clip,keepaspectratio]{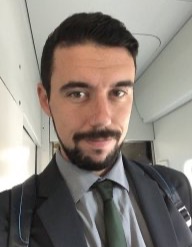}}]{Matteo Saviozzi}
(Senior Member, IEEE) received the M.Sc. degree in mathematics and the Ph.D. degree in electrical engineering from the University of Genoa, Genoa, Italy, in 2012 and 2017, respectively. He is currently an Associate Professor with the Department of Electrical, Electronics, Telecommunications Engineering, and Naval Architecture (DITEN), University of Genoa. His research focuses on the study, implementation, and impact of distributed generation from renewable and conventional sources in the distribution network, and energy consumption management. His key interests include optimization algorithms, distribution management systems, load and renewable energy resource forecasting, load modeling, energy management, artificial intelligence methods, and probabilistic security assessment. He is currently a member of CIGRE's Working Group C6/C2.34 and the Technical Committee on Power and Energy Systems of IFAC.
\end{IEEEbiography}
\begin{IEEEbiography}[{\includegraphics[width=1in,height=1.25in,clip,keepaspectratio]{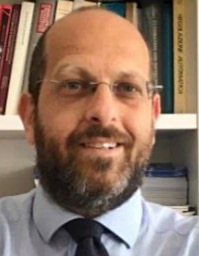}}]{Federico Silvestro}
(Senior Member, IEEE) was born in Genoa, Italy, in 1973. He received the Electrical Engineering degree and the Ph.D. degree in electric power systems from the University of Genoa, Genoa, Italy, in 1998 and 2002, respectively. He is currently a Full Professor with the Department of Electrical, Electronics and Telecommunication Engineering and Naval Architecture, University of Genoa, and the Deputy Chair of the Department. He has been scientific responsible for several national and international research projects and has authored more than 250 scientific papers. His research interests include power system optimization and microgrids, flexibility and multi-energy systems, marine applications, and knowledge-based systems applied to power systems. He is the Secretary of the CIGRÉ Working Group JWG C1/C6.37/CIRED on Optimal Transmission and Distribution Investment Decisions under Increasing Energy Scenario Uncertainty.
\end{IEEEbiography}
\begin{IEEEbiography}[{\includegraphics[width=1in,height=1.25in,clip,keepaspectratio]{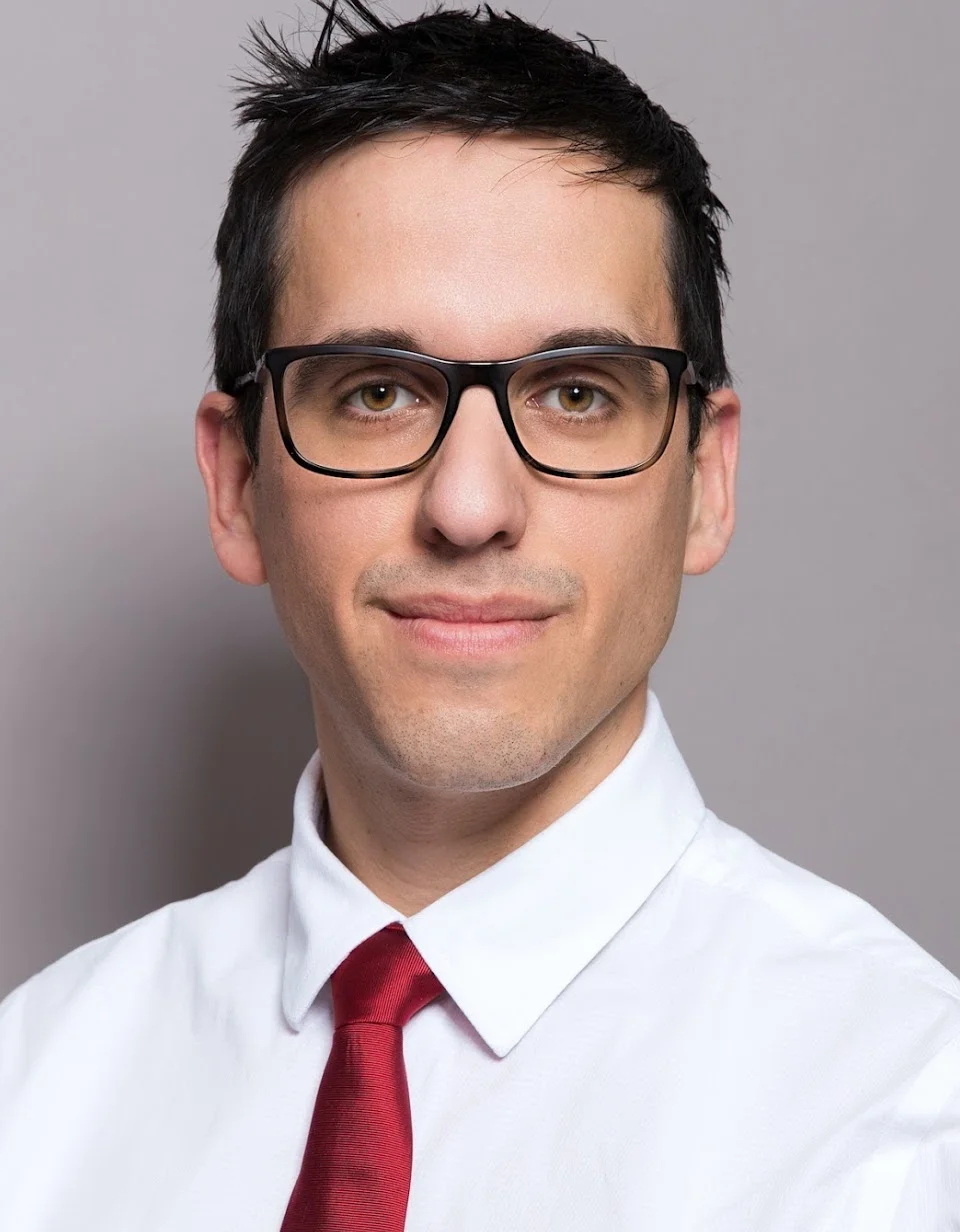}}]{Fabrizio Sossan}
is a professor of electrical power systems with the University of Applied Sciences of Western Switzerland (HES-SO Valais-Wallis). His research focuses on planning, scheduling, and control for energy storage and renewables in power grids. He leads Gridlab at HES-SO Valais-Wallis. He received the Ph.D. degree in electrical engineering from the Technical University of Denmark (DTU) in 2014 and the master's degree in computer engineering from the University of Genova, in 2010.
\end{IEEEbiography}
\end{document}